\documentclass[11pt]{article}
\usepackage{import}
\usepackage{xifthen}
\usepackage{pdfpages}
\usepackage{transparent}

\usepackage[english]{babel}
\usepackage[colorinlistoftodos]{todonotes}
\usepackage{xcolor}
\usepackage{listings}
\usepackage{hyperref}
\hypersetup{
    colorlinks=true,
    linkcolor=blue,
    filecolor=purple,      
    urlcolor=cyan}
\usepackage{wrapfig}
\usepackage[utf8]{inputenc}
\usepackage[T1]{fontenc} 
\usepackage[backend=biber,style=ieee,sorting=none]{biblatex} 
\usepackage{textcomp}
\usepackage{csquotes}
\usepackage[nottoc,numbib]{tocbibind} 
\usepackage{pdfpages}
\usepackage[margin=1.1in]{geometry} 
\usepackage{amsmath}
\usepackage{graphicx}
\usepackage{epstopdf}
\graphicspath{ {Images/} } 
\usepackage[export]{adjustbox}
\usepackage{subcaption}
\usepackage{wrapfig}
\usepackage{rotating}
\usepackage{amssymb}
\usepackage{physics}
\usepackage{multicol} 
\usepackage{fancyhdr} 
\usepackage{float}
\usepackage{import}
\usepackage{enumerate}
\usepackage{enumitem}
\usepackage{parskip}
\usepackage[format=plain, font=it]{caption}
\usepackage{siunitx}
\usepackage{chngcntr}
\counterwithin{figure}{section}
\counterwithin{table}{section}
\usepackage{listings} 
\usepackage{titlesec}
\titleformat{\paragraph}
{\normalfont\normalsize\bfseries}{\theparagraph}{1em}{}
\titlespacing*{\paragraph}
{0pt}{3.25ex plus 1ex minus .2ex}{1.5ex plus .2ex}
\begin{document}

\begin{titlepage}
\newcommand{\HRule}{\rule{\linewidth}{0.5mm}} 

\begin{center} 
 




{\huge Dynamics of ion temperature gradient modes in burning plasma conditions in the presence of energetic particles}
\\[0.5cm]

{R. Ivanov$^{1,2*}$, A. Biancalani$^2$, A. Bottino$^3$, D. Gossard$^2$, T. Hayward-Schneider$^3$,A. Mishchenko$^4$, and R. Wu$^{2}$}
\\[0.5cm]
{$^1$Laboratoire de Physique des Plasmas, CNRS, Université Paris Saclay, Ecole Polytechnique, Sorbonne Université, Observatoire de Paris, F-91120 Palaiseau, France}

{$^2$De Vinci Higher Education, De Vinci Research Center, 92916 Paris, France}

{$^3$ Max Planck Institut für Plasmaphysik, Boltzmanstrasse 2, D-85748 Garching, Germany}

{$^4$ Max Planck Institut für Plasmaphysik, Wendelsteinstraße 1, D-17491 Greifswald, Germany}
\\[0.5cm]
{$^*$Author to whom any correspondence should be addressed.}

{\textbf{E-mail:} roman.ivanov@lpp.polytechnique.fr}

{\textbf{Keywords:} Energetic particles, gyrokinetics, ion temperature gradient modes, dilution}\\[1cm]

{\Large\bfseries Abstract}
\end{center}
The interaction between energetic particles (EPs) and ion temperature gradient (ITG) modes is studied using the global particle in cell ORB5 code. In this work, we extend previous studies to a broader range of EP temperatures, including the burning plasma regime and to wider variety of EP distribution functions. Two main stabilization mechanisms are found to be effective in ITG stabilization confirming previous studies: direct dispersion relation modification (DDRM) effective only at intermediate EP temperatures and dilution effect (DE) which is independent of EP temperature and becomes dominant in burning plasma regime ($T_f > 50\,T_i$). The study is further extended to slowing-down EP distributions which in contrast exhibit no DDRM-related stabilization. 
The findings are further validated in an ITER pre-fusion operation scenario and additionally compared with electromagnetic effects. In this scenario EP stabilization is found to be weaker than $\beta$-stabilization. Overall, these results provide better understanding of EP–ITG interactions over a wider range of EP parameters relevant to burning plasma regime which is important for predicting turbulence and confinement in future devices such as ITER.\\[1cm]

\begin{center}
{\today} 
\end{center}

\vfill 

\end{titlepage}

\newpage

\setlength{\parskip}{\baselineskip}

\section{Introduction} \label{introduction}
Fusion holds the promise of clean and virtually limitless energy production on Earth. Achieving controlled fusion reactions, however, requires a deep understanding of plasma physics and the ability to accurately diagnose and control the behavior of plasma under extreme conditions.

One of the key problems of modern tokamaks and future fusion reactors is turbulence and the resulting anomalous transport. There are many types of turbulence that arise under different conditions and in different regions of the plasma, and it is impossible to avoid all of them in a single configuration. They tend to transport particles and heat in an unfavorable direction, toward the wall, thus degrading the confinement and possibly leading to the breakdowns.  One of the most significant is an ion temperature gradient (ITG) turbulence \cite{Rudakov1961}\cite{Coppi1967}, that arises in the presence of the temperature gradient in the curved magnetic field line configuration as is the case in tokamaks. 

Turbulent modes have different stabilization mechanisms, one branch being a stabilization by the energetic particles (EPs), which was seen in the experiments and modelling of JET tokamak \cite{Romanelli2010}\cite{Bonanomi2018}. In reactor-scale plasma energetic particles are inevitably generated via two mechanisms: fusion reactions and auxiliary heating (neutral beam injection (NBI) and radio frequency (RF) heating: ion cyclotron resonance heating (ICRH) and lower hybrid heating (LHH)). They have temperatures several times exceeding the temperatures of thermal ions and can have different interaction mechanisms with ITG.

Mechanisms of the EP stabilization of the micro-turbulence \cite{Citrin2023} were studied with global codes (ORB5 \cite{Lanti2020}, GENE \cite{Gorler2011}). Recent studies show that two main electrostatic mechanisms of stabilization are found to be efficient: direct dispersion relation modification (DDRM) \cite{DiSiena2018}\cite{DiSiena2025} effective at intermediate EP temperatures for Maxwellian EPs and dilution effect (DE) \cite{Tardini2007},\cite{Wilkie2017},\cite{Wilkie2018}\cite{Citrin2013}. The latter is expected to dominate in the burning plasma regimes, although no comprehensive studies have been done so far. In these studies it was shown that under certain conditions the stabilization by the EPs can be favored. However, parameter space governing these effects and their influence is not fully investigated and explained. 

Present work extends previous analyses on investigating the impact of EPs on ITG turbulence using ORB5 simulations in the burning plasma regime. Maxwellian and slowing-down EP distributions are considered, focusing on understanding of the role of EP temperature, density, and anisotropy. The study confirms two mechanisms of ITG stabilization: the direct dispersion relation modification (DDRM) and the dilution effect (DE), and examines their relevance under simplified analytical case and experimental ITER case in comparison with electromagnetic stabilization effects.

The paper is structured as follows. An introduction to the physics of the reduction mechanisms of the linear ITG growth is given in Sec. \ref{sec:itg}. The competition of these mechanisms in the electrostatic limit, and for a regime going from present tokamak plasmas to burning plasmas, is studied by means of global simulations, and described in Sec. \ref{sec:results}. Moreover, in the same section, the importance of the beta stabilization in burning plasmas is also stressed by means of dedicated electromagnetic simulations. Finally, a discussion is presented in Sec. \ref{conclusion}, where the summary of the results is also presented.

\section{ITG stabilization mechanisms}\label{sec:itg}
\subsection{Direct effect of the EPs on the ITG modes}
Energetic particles (EPs) can interact with ITG and modify directly its dispersion relation. This direct effect of the EPs on the ITG was initially described in \cite{DiSiena2018}, \cite{DiSiena2025} under the name of "resonance effect". To be precise, due to the finite orbit width effect, a clean resonance is difficult to occur here, especially when we investigate large EP temperature, where the orbit width becomes very large. A detailed physical interpretation of such localized interactions was studied for the geodesic acoustic modes (GAMs) \cite{Qiu2009}, \cite{Chen2018}. Although, ITG modes differ from GAMs in frequency and spatial structure, a similar mathematical mechanism underlies the EP contribution to the ITG response. Generally speaking, we refer to this mechanism as "direct dispersion relation modification (DDRM) 

According to the reduced collision-less Vlasov ES model without trapping, effective interaction between the particles and the wave occurs when several conditions are satisfied. The perturbed distribution function of EPs eq. \ref{eq:F1_DiSiena}, derived in the paper \cite{DiSiena2018}, has a dominant contribution in the dispersion relation when the denominator vanishes:
\begin{equation}\label{eq:F1_DiSiena}
    F_1 = \frac{
    k_y \phi_1 \left[ \frac{1}{2v_{\parallel}} \frac{\partial F_0}{\partial v_{\parallel}} \left( \frac{\mu B_0 + 2v_{\parallel}^2}{B_0} \right) \mathcal{K}_y - \frac{1}{C} \hat{\partial}_x F_0 \right]
}{
    \omega_r + i\omega_i + \frac{T_f}{q} \left( \frac{\mu B_0 + 2v_{\parallel}^2}{B_0} \right) k_y \mathcal{K}_y
},
\end{equation}
where $\phi_1$ is perturbed scalar potential, $w_r$ and $w_i$ - are real and imaginary parts of the mode pulsation, $k_y$ - is a mode y-wavenumber, $q$ - safety factor, $B_0$ - on-axis magnetic field, $\mu$ and $v_\parallel$ are velocity space coordinates normalized to the $(2T_f/m_f)^{1/2}$ and $T_f/B_0$ respectively. Last term in the numerator for the Maxwellian distribution $F_0$ of EPs is the following:
\begin{equation}\label{eq:dF_DiSiena}
- \hat{\partial}_x F_{0,M} = \left[ \frac{R}{L_n} + \frac{R}{L_T} \left( v_{\parallel}^2 + \mu B_0 - \frac{3}{2} \right) \right] F_{0,M}
\end{equation}

DDRM condition from (eq. \ref{eq:F1_DiSiena}) can be satisfied only in the low-field side of the tokamak, where the magnetic curvature $\mathcal{K}_y$ is negative. In order for the EPs to have stabilizing effect, condition on $\eta_f^{-1}=L_{T,f}/L_{n,f}<1.5$ should be fulfilled, which is derived from the eq. \ref{eq:dF_DiSiena}.

For the Maxwellian distribution, there are three regimes of different $T_f$ to be discussed for DDRM: low $T_f<6\,T_i$, high $T_f>30\,T_i$ and intermediate region for the values of $T_f$ in between.
In the denominator of eq. \ref{eq:F1_DiSiena} quantities $q$, $\omega$ and $k_y$ are preserved, by changing $T_f$ we change only part of the velocity space that fulfill the zero of the denominator. For low $T_f$ velocities fulfilling the DDRM conditions are relatively large, from the tails of the Maxwell distribution, so the fraction of the particles that is interacting with the ITG via the DDRM is negligible. Thus, wave-fast ion interaction is only shaped by the nominator and the drive term $\hat{\partial}_x F_{0,M}$, because in this regime $\mathcal{K}_y$ is  close to $0$ as was shown in \cite{DiSiena2018}. In the region of high temperatures, the wave-fast ion interaction almost disappears (see sec. \ref{sec:DE}), so DDRM there is also not effective.

On the contrary, in the region of intermediate temperatures DDRM equation can be fulfilled as there is a significant fraction of fast ions for the ITG to interact with. Once it is fulfilled, the drive term defines the quantity and sign of the growth rate: as far as it is proportional to $(v^2_\parallel+\mu B_0-3/2)\cdot F_{0,M}(v_\parallel,\mu)$ it can be positive or negative and has clear extreme points. When it is negative, $\gamma$ is negative - stabilization, opposite - destabilization. In its minimum negative value $\gamma$ is the most negative, thus the stabilization is the most efficient. As far as the dependency of the drive term on the $(v^2_\parallel+\mu B_0)$ has a local minimum, the growth rate should also have a clear minimum in the $T_f$ dependency.

\subsection{Dilution effect}\label{sec:DE}

Dilution effect (DE) was described in the paper \cite{Tardini2007} that also confirmed this effect in the experimental case of ASDEX Upgrade. This effect was also depicted algebraically in paper \cite{Wilkie2017}.

It can be shown that the EP growth rate contribution:
\[
\gamma_f = -\frac{1}{E_{\mathrm{pot}}}
\left\{
\int dz\, dv_{\|}\, d\mu\, \pi n_f J_0^2 \, |\phi_1^{k}|^2 T_f \left( 2 v_{\|}^2 + \mu B_0 \right)
\, k_y^2 \mathcal{K}_y \frac{\gamma}{C}
\frac{\left[ \frac{R}{L_n} + \frac{R}{L_T} \left( v_{\parallel}^2 + \mu B_0 - \frac{3}{2} \right) \right] F_{0,M}
}{
    (\omega_r + \frac{T_f}{q} \left( \frac{\mu B_0 + 2v_{\parallel}^2}{B_0} \right) k_y \mathcal{K}_y)^2+\gamma^2
}
\right\}
\]

in the limit of high $T_f$ goes to $0$. Thus, EPs at high temperatures (compared to the background ions: $T_f> 50\,T_{i}$) interact weakly with ITG, substituting the fraction of ions that had strong interaction with it. 

In the slab ITG dispersion relation, and likewise for the toroidal ITG mode, the linear growth rate is proportional to the factor $n_i/n_e$. In a plasma without impurities or fast ions this ratio is unity due to quasineutrality, and therefore does not explicitly appear.

However, in the dilution regime the presence of an energetic-particle population modifies the ion density, giving an effective factor $1-n_f/n_e$. Consequently, the DE stabilizes the ITG mode through the direct reduction of this proportionality factor. At sufficiently high energetic-particle temperatures, DE becomes the dominant stabilization mechanism, while the DDRM becomes negligible in this regime.

\subsection{EP effect on ITG in the slowing-down distribution case}\label{sec:SD}
Energetic particles in tokamak plasma have more sophisticated distributions than Maxwellian, which is still can be used as a simple approximation. More precise approximation with analytical functions can be done by the slowing-down distribution (SD). For the fusion products an appropriate distribution is an isotropic SD, which is described by the following equation: 
\begin{equation}
F_{0,s} = 
\frac{n_{0}}{\frac{4}{3}\pi \log\!\left[ 1 + \left( \frac{v_{f}}{v_{c}} \right)^{3} \right]}
\frac{\Theta(v_{f} - v)}{v_{c}^{3} + v^{3}},
\end{equation}

where $v_{f}$ is 
$v_{f} = \sqrt{2E_{f}/m_{f}}$, $E_{f}$ is the birth energy,
$
v_{c} = \sqrt{\frac{2T_{e}}{m_{e}}}
\left( \frac{3\sqrt{\pi}}{4} 
\sum_{\mathrm{ions}} \frac{n_{i} m_{e} q_{i}^{2}}{n_{e} m_{i}} \right)^{1/3}
$
and $\Theta$ is a Heaviside function. Anisotropic distribution can describe well the neutral beam injected ions. Analytical expression for it is shown in paper \cite{Rettino2022} and has a sophisticated form that adds two main parameters that can change the form of the distribution: $\xi=v_\parallel/\vert v\vert$ - is an anisotropy parameter that defines the angle of injection and $\sigma$ is a parallel velocity dispersion parameter. For both isotropic and anisotropic cases the temperature parameter is relevant. But in contrast to the Maxwellian case, temperature of the SD defines the birth energy of the EPs, so in order to compare the temperature scans the equivalent to Maxwellian temperature needs to be defined for the SD: \begin{equation}\label{eq:t_equiv}
    T_{f} = \frac{2}{3} \frac{I_4\left( \frac{v_c}{v_{f}} \right)}{I_2\left( \frac{v_c}{v_{f}} \right)} \mathcal{E}_{f},
\end{equation}
where $I_n(a)=\int_{0}^{1} x^n/(x^3+a^3) \,dx$. In the following sections for the SD distribution the Maxwellian temperature will be used if not stated the opposite.

In the case of the SD distribution, the drive term can be expressed as following \cite{Vanini2023}:
\begin{equation}
\hat{\partial}_x F_0= 
\left\{
\hat{\partial}_x \ln n_{f}
+ \left(
\frac{\left(v_{f}/{v_c}\right)^3}{\left[1+\left(v_{f}/{v_c}\right)^3\right]\ln\!\left[1+\left(v_{f}/{v_c}\right)^3\right]}
- \frac{v_c^3}{v_c^3+v^3}
\right) \cdot 3 \, \hat{\partial}_x \ln v_c
\right\} F_0
\end{equation}

This term, as in a Maxwellian case, can have both signs depending on the plasma profiles and velocities. Depending on $v$, drive term can change sign, thus the growth rate contribution from the EP fraction can vary. Similarly to the Maxwellian case, the criterion on the destabilization can be deduced for the isotropic SD case, by taking the value of $v$ equal to $0$. In this case,  the criterion can be written as the following, exhibiting the destabilization in case:
$$
\frac{L_{T,e}}{L_{n,f}} > \frac{3}{2}\cdot \left(1-\frac{\left(v_{f}/{v_c}\right)^3}{\left[1+\left(v_{f}/{v_c}\right)^3\right]\ln\!\left[1+\left(v_{f}/{v_c}\right)^3\right]}\right)
$$

\subsection{Electromagnetic stabilization of the ITG}
Electromagnetic stabilization of the ITG mode was observed and described previously in papers \cite{Kim1993}, \cite{Hong1989}. It was shown that electromagnetic (finite-$\beta$) effects reduce ITG growth by coupling the ITG branch to the shear-Alfven response even in the absence of the EP population. In the pure thermal-ion case, ITG stabilization threshold is found to be also the destabilization threshold for the kinetic ballooning mode (KBM) \cite{Zonca1999}. As a consequence, the ITG growth rate decreases with increasing $\beta$ up to this threshold, beyond which the KBM becomes destabilized and the total growth rate increases again with stronger dependency. EM stabilization is not the main subject of this paper, being present only in the last section \ref{sec:ITER}.
\section{Results}\label{sec:results}
\subsection{Ad-hoc simplified case} 

In this work we used two different cases of geometry and equilibrium to study. The first case is taken from the paper \cite{Mishchenko2022}. Tokamak geometry is chosen to have circular concentric flux surfaces with a simplified magnetic equilibrium defined by these equations:
\begin{equation}
    \mathbf{B}=\nabla \psi \times \nabla \varphi + I \nabla \varphi \quad \mathrm{with} \quad I = B_0 R_0, \quad \psi(\rho) = \int_0^{\rho} \frac{B_0 \rho' \, d\rho'}{q(\rho')}
\end{equation}
where $B_0=1\,T$ - an on-axis magnetic field, $R_0=10\,m$ - major radius, $a=1\,m$ - minor radius. Safety factor profile is chosen to be $q(\rho)=1.1+0.8\rho^2$, where $\rho$ is a radius of the circular flux surface. Density and temperature profiles are defined by following equations:
\begin{equation}
    n_{0s}(s)/n_{0s}(s_0)= \exp \left[ -\kappa_n \Delta_n \tanh \left( \frac{s - s_0}{\Delta_n} \right) \right]
\end{equation}

\begin{equation}
    T_{0s}(s)/T_{0s}(s_0) = \exp \left[ -\kappa_T \Delta_T \tanh \left( \frac{s - s_0}{\Delta_T} \right) \right]
\end{equation}
where $s=\sqrt(\psi/\psi_a)$, $s_0=0.5$, $\Delta_T=\Delta_n=0.208$ and $\kappa_T=2$, $\kappa_n=0.3$ for all the simulations except the case with the Maxwellian destabilization EPs, where $\kappa_T=0.5$, $\kappa_n=2$ were used.
Simulations had three species: adiabatic electrons, kinetic hydrogen thermal ions and kinetic hydrogen energetic particles (EPs). Temperature of electrons and hydrogen were the same, while for the EPs it was scanned in the interval $T_f=1\dots100$ last corresponding to the regimes of burning plasma. Concentration of the EPs was taken $n_f/n_{i}=10\%$ for most of the studies, except the studies, where it was changed to keep $\beta_f$ value constant with the temperature scan $T_f=1\dots100$.

\subsubsection{Maxwellian distribution of energetic particles} 
\begin{figure}[h]
\begin{center}
\begin{minipage}[h]{0.49\linewidth}
\includegraphics[width=1\linewidth]{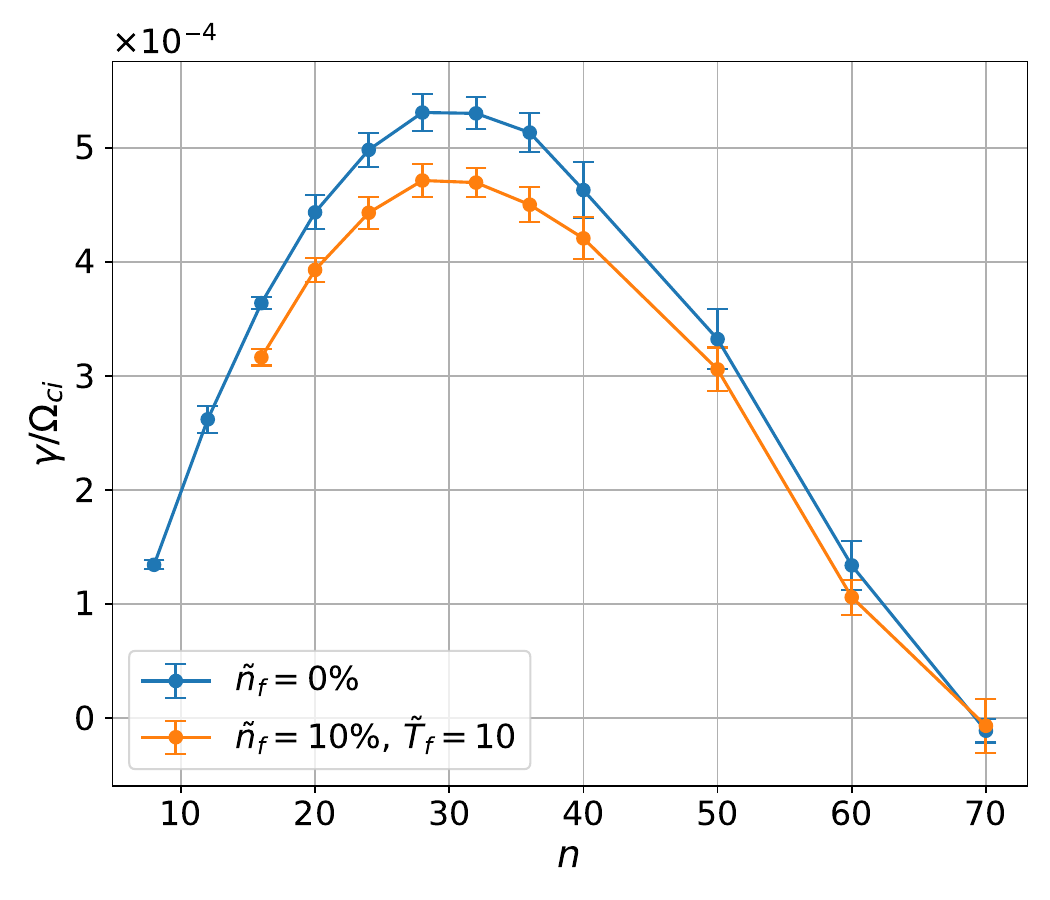}
\end{minipage}
\hfill
\begin{minipage}[h]{0.49\linewidth}
\includegraphics[width=1\linewidth]{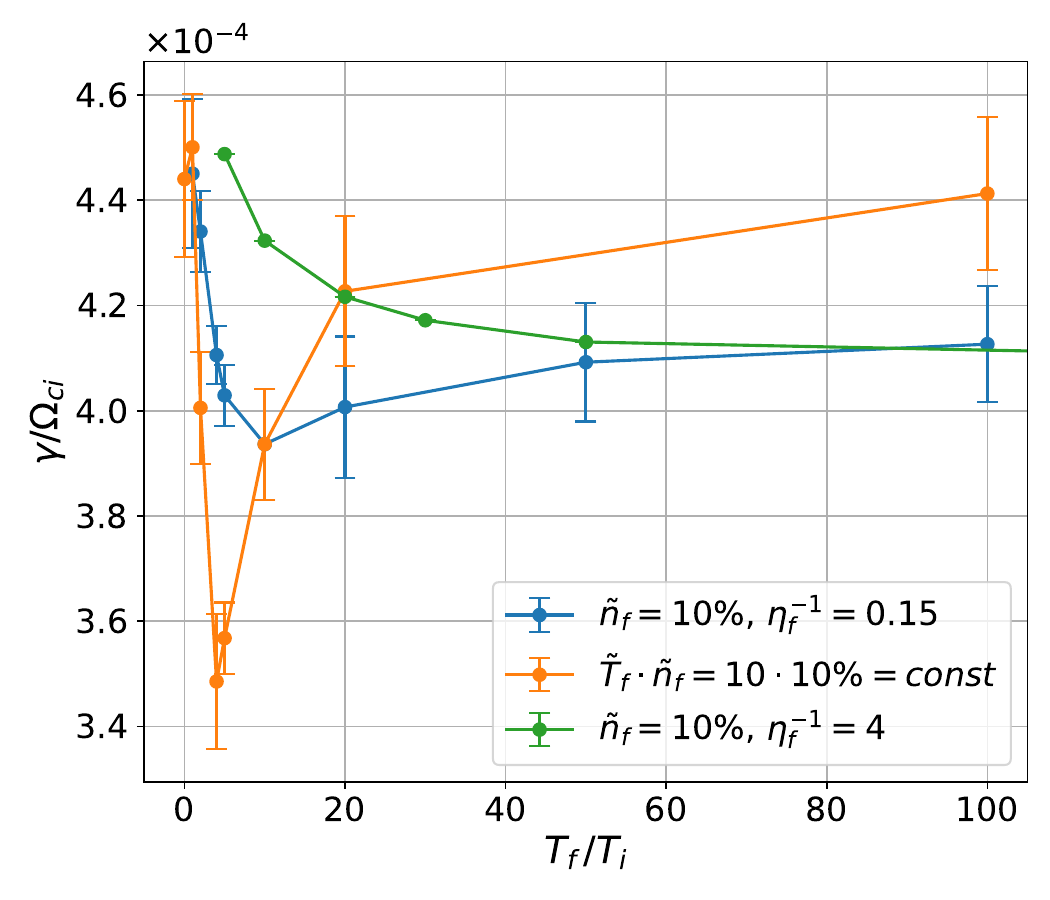}
\end{minipage}
\end{center}
\caption{a) - growth rate $\gamma(n)$ scan in single toroidal mode number for two cases: without EPs and with EPs of concentration $n_f/n_i=0.1$ and temperature $T_f/T_i=10$. Presence of EPs only suppresses $\gamma (n) $ without changing the spectrum shape; b) - growth rate $\gamma(n)$ scan in EP temperature for constant EP density in stabilizing regime $\eta_f^{-1}=0.15$ (blue) and destabilizing $\eta_f^{-1}=4$ (green), constant EP beta (yellow). DDRM effect is observed at $T_f\approx10\,T_i$ for a blue curve, while the steeper drop is seen for the yellow curve for lower temperatures due to the higher concentration of stabilizing EPs, for the green curve there is no DDRM as the condition on zero of the denominator is not fulfilled}
\label{fig:gamma_n}
\end{figure}

Linear ITG instability was at first studied in the single toroidal mode number regime in a simplified "ad-hoc" case and in the range $m=-128\dots 128$ of poloidal number. The dominating poloidal mode was found to be $m=27$ at the radial location $s=0.46$. The growth rate of the mode was extracted from the slope of the logarithmic scalar potential evolution at this location. First, the toroidal mode number scans were performed for cases without and with energetic particles with temperatures of $T_f/T_i=10$ and Maxwellian distribution. In the Fig. \ref{fig:gamma_n} these scans are compared showing peaked behavior at the toroidal mode numbers $n=28\dots 32$, the only difference is in the reduction of the growth rate for the case with EPs without changing the shape of a curve. 

Further, the growth rate temperature scan was performed for the fixed toroidal mode number $n=20$ for three cases: constant EP density $n_f/n_i=10\%$ with parameter $\eta_f^{-1}=0.15$ which corresponds to the stabilizing regime, same scan but with destabilizing $\eta_f^{-1}=4$ and with constant EP beta $\beta_f$ kept the same as at the point $n_f/n_i=10\%$ and $T_f/T_i=10$ in the stabilizing regime $\eta_f^{-1}=0.15$. First two scans were intended to confirm the results of the paper \cite{DiSiena2018} and to expand it to the burning plasma regime, while the second is more relevant for fusion plasma, where beta parameter is kept constant. These scans are compared in the Fig. \ref{fig:gamma_n}. 

Constant density scan with $\eta_f^{-1}=0.15$ represents the same behavior observed before: low EP temperatures have comparable EP contributions to the ITG, whereas at the intermediate temperatures $T_f\approx10$ the DDRM is strong and the "well-like" shape is present. At higher temperatures, the EP contribution starts to diminish and reaches the value of pure dilution effect, at burning regime with $T_f /T_i = 100$, consistently with the conjecture of \cite{DiSiena2018}. Constant density scan with $\eta_f^{-1}=4$ on the contrary has no DDRM stabilization, as $\eta_f$ parameter is set in a way that drive term is always negative. Thus we observe the monotonous decrease in the growth rate with temperature, as it is approaching the DE.

Constant beta scan has different behavior: it has a shifted minimum to the temperatures of $T_f\approx4\,T_i$, which can be explained by the fact that at this point EP density is relatively high $n_f/n_i=20\%$ and EP contribution continues to grow in absolute value with temperature. If the temperature is decreased, the overall growth rate starts to grow because the EP gamma contribution becomes positive and starts to have a destabilizing effect. The increase in EP density thus makes this dependence more strong which is observed with a sharper slope. At higher temperatures, where DDRM is reduced, the DE starts to dominate and the overall growth rate decreases linearly with temperature, as density is decreased to keep $\beta_f=const$.

\begin{figure}[h]
\begin{center}
\begin{minipage}[h]{0.45\linewidth}
\includegraphics[width=1\linewidth]{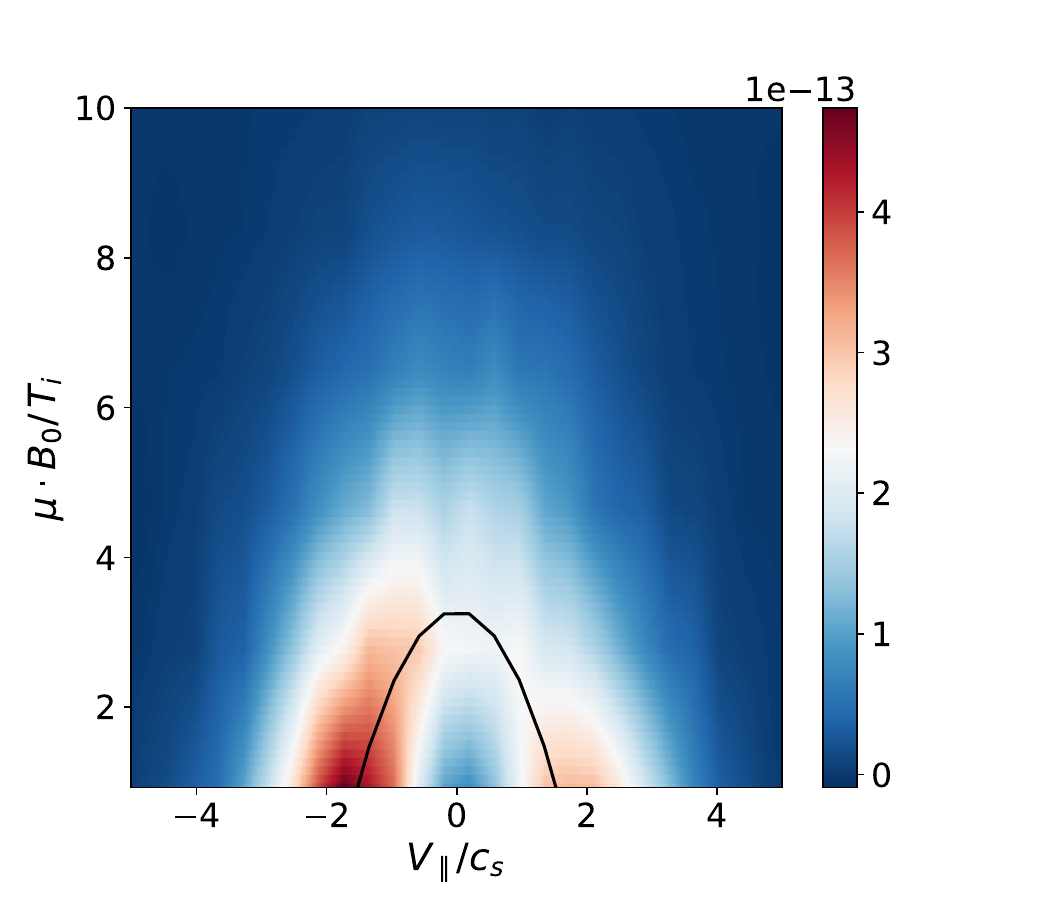}
\end{minipage}
\hfill
\begin{minipage}[h]{0.49\linewidth}
\includegraphics[width=1\linewidth]{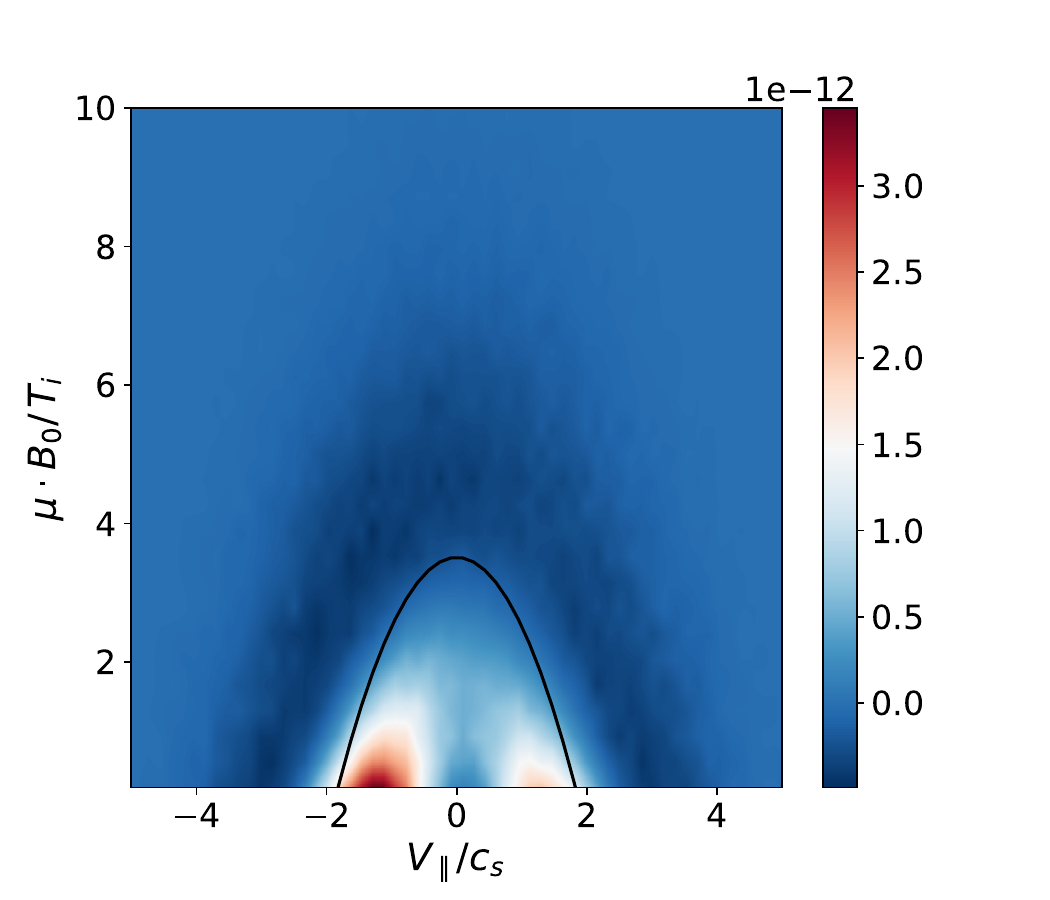}
\end{minipage}
\end{center}

\caption{Power exchange between the Maxwellian EPs and ITG from the MPR diagnostic, at different temperatures: a) - $T_f=10\,T_i$. Power exchange is effective in the parabola region signifying an effective DDRM; b) - $T_f=2\,T_i$, power exchange is effective in the region where the most particles are, without effective interaction in the parabola region, DDRM is ineffective. Black parabola corresponds to zero of the denominator of eq. \ref{eq:F1_DiSiena}.}
\label{fig:MPR}
\end{figure}

Mode particle resonance (MPR) diagnostic \cite{Novikau2021} was used to compare the phase space domains of the EP distribution that interact strongly with the ITG. In Fig. \ref{fig:MPR} two MPR heat-maps showing the power exchange between the ITG and EPs are compared for Maxwellian EPs with temperatures of $2$ and $10\,T_i$. The curve corresponding to the zero of the denominator of the eq. \ref{eq:F1_DiSiena} is plotted on top. For the $10\,T_i$ case the shape of the power exchange coincides with the parabola confirming again an effective DDRM, while for the $2\,T_i$ case the interaction is the most effective in the region where the most particles are distributed without any increase in the region of the zero of the denominator.

\subsubsection{Slowing-down distribution of energetic particles} \label{SD}

\begin{figure}[h]
\begin{center}
\begin{minipage}[h]{0.49\linewidth}
\includegraphics[width=1\linewidth]{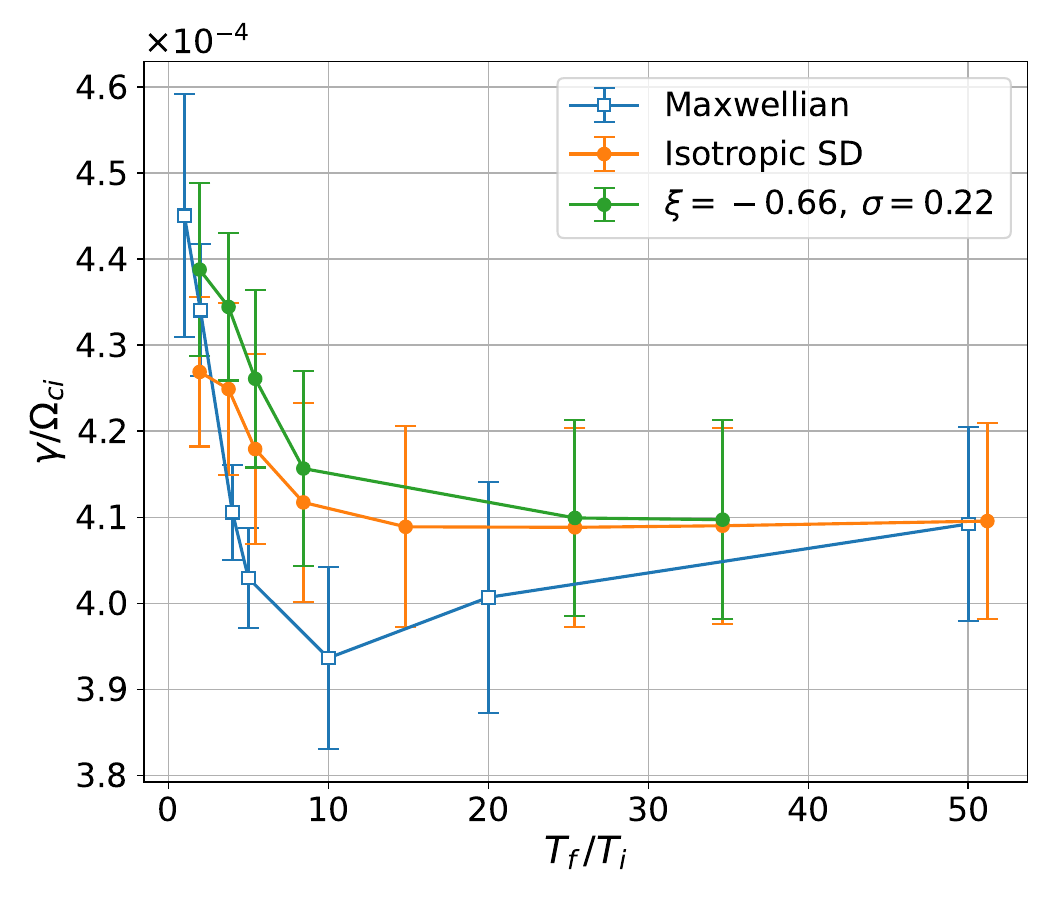}
\end{minipage}
\hfill
\begin{minipage}[h]{0.49\linewidth}
\includegraphics[width=1\linewidth]{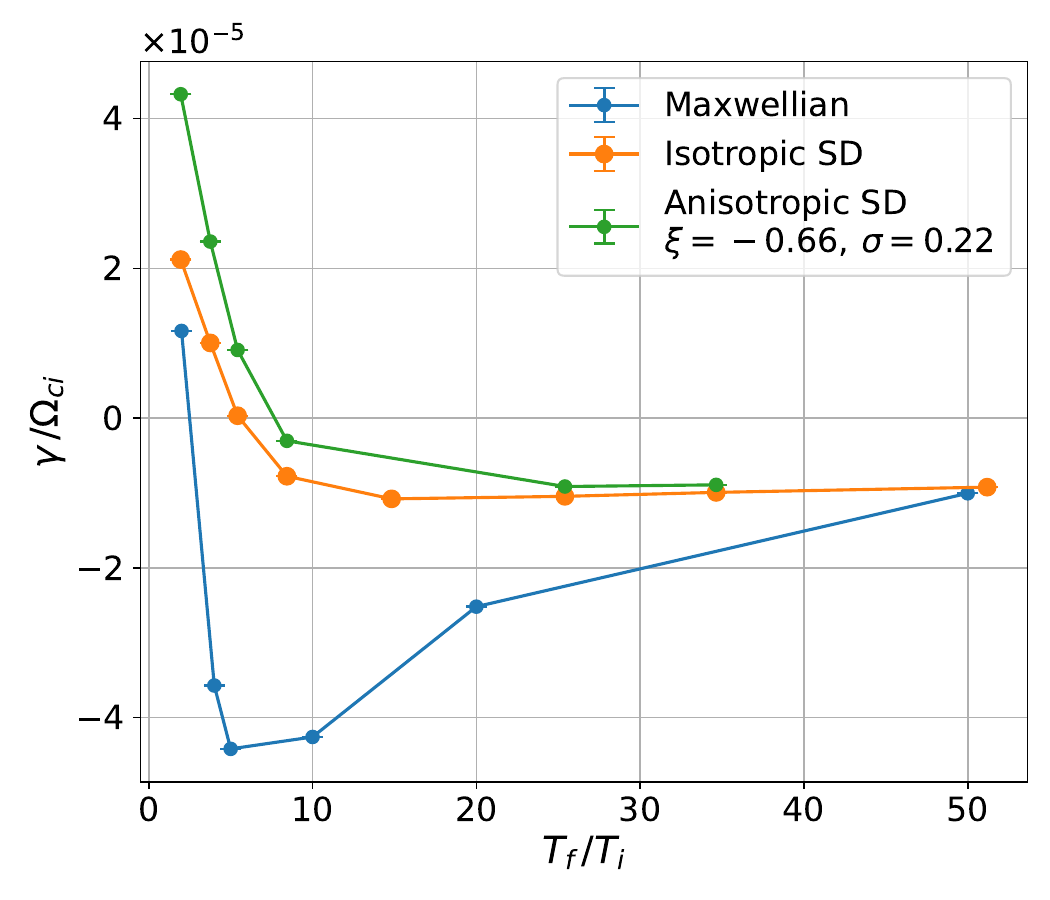}
\end{minipage}
\end{center}

\caption{a) - growth rate scan in EP temperature from the scalar potential evolution for constant EP density for different EP distributions: Maxwellian, isotropic SD, anisotropic SD with $\xi=-0.66$ and $\sigma=0.22$; b) - EP growth rate scan from the MPR diagnostic DDRM is observed only in the Maxwellian case, while in SD the behavior is monotonous with the EP destabilization as the drive term is always negative}
\label{fig:gamma_SD}
\end{figure}

Growth rate temperature scans were performed for SD distributions both isotropic and anisotropic with $\xi=-0.66$ and $\sigma=0.22$ and compared with Maxwellian in Fig. \ref{fig:gamma_SD}. The SD have different behavior from the Maxwellian case: it is monotonously decreasing with the temperature which signalizes that the DDRM in this case is inefficient. This behavior is similar to the one of Maxwellian with $\eta_f^{-1}=4$ shown before.

\begin{figure}[h]
\begin{center}
\begin{minipage}[h]{0.49\linewidth}
\includegraphics[width=1\linewidth]{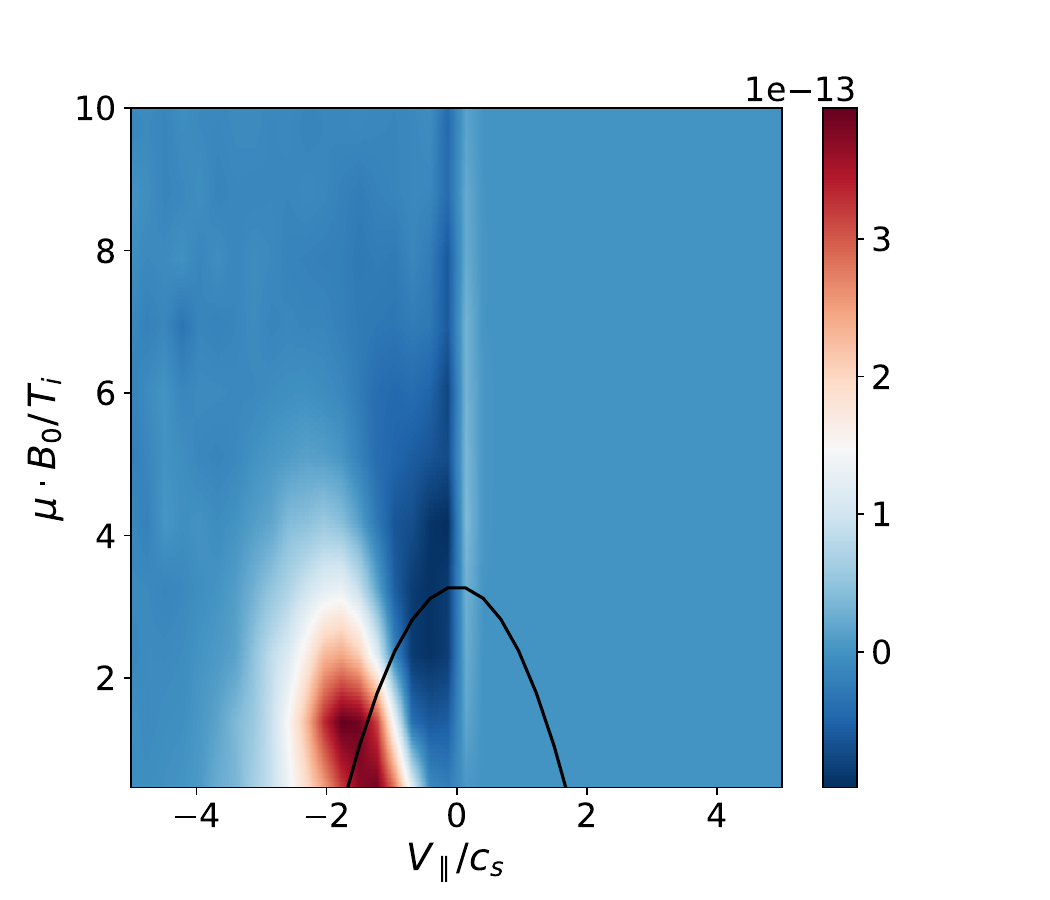}
\end{minipage}

\caption{Power exchange between EPs and ITG from the MPR diagnostic in the adhoc case with anisotropic SD EPs with $\xi=-0.66$ and $\sigma=0.22$: DDRM is ineffective, as power exchange occurs not at DDRM condition (black)}
\label{fig:MPR_SD}
\end{center}
\end{figure}

MPR diagnostic was also used in this anisotropic SD case Fig.\ref{fig:MPR_SD}. As in "low-temperature" Maxwellian case, interaction occurs in the region where the most particles are distributed without any increase in the region of the DDRM. In Fig. \ref{fig:gamma_SD} the growth rate contribution from the EPs only extracted from the MPR is shown. Effective DDRM for the Maxwellian as well as its absence for the SD is clearly seen.

For the isotropic SD case the transition from destabilization to stabilization is seen for temperatures in range $4\dots7\,T_i$. This occurs due to a change in the sign of the drive term. For it the criterion was derived in the Sec. \ref{sec:SD}. In this set of simulations, this criterion gives an effective switch of the sign for the equivalent Maxwellian temperature $T_f=2\,T_i$, while the transition in the Fig. \ref{fig:gamma_SD} is seen at higher temperature. This result is still very close and can be seen as a numerical uncertainty.

\clearpage
\newpage

\subsection{ITER scenario} \label{sec:ITER}
Previous sections were devoted to the simplified case that allows to study the EP ITG interaction varying many parameters and having many computationally cheap simulations. This section describes the results of the ITER case. Due to the more complex magnetic geometry and the larger ratio of tokamak spatial scales vs kinetic scales, these simulations are more numerically demanding. Thus, the set of diagnostics as well as the range of parameters varied is smaller. Experimental scenario of ITER PFPO 101006 was considered, which was shown in the paper \cite{Hayward2022}. Main parameters are: magnetic field $B_0=2.65\, T$ on axis, major radius is $R=6.2\,m$, minor: $a=2\,m$. 

\begin{figure}[h]
\begin{center}
\begin{minipage}[h]{0.49\linewidth}
\includegraphics[width=\linewidth]{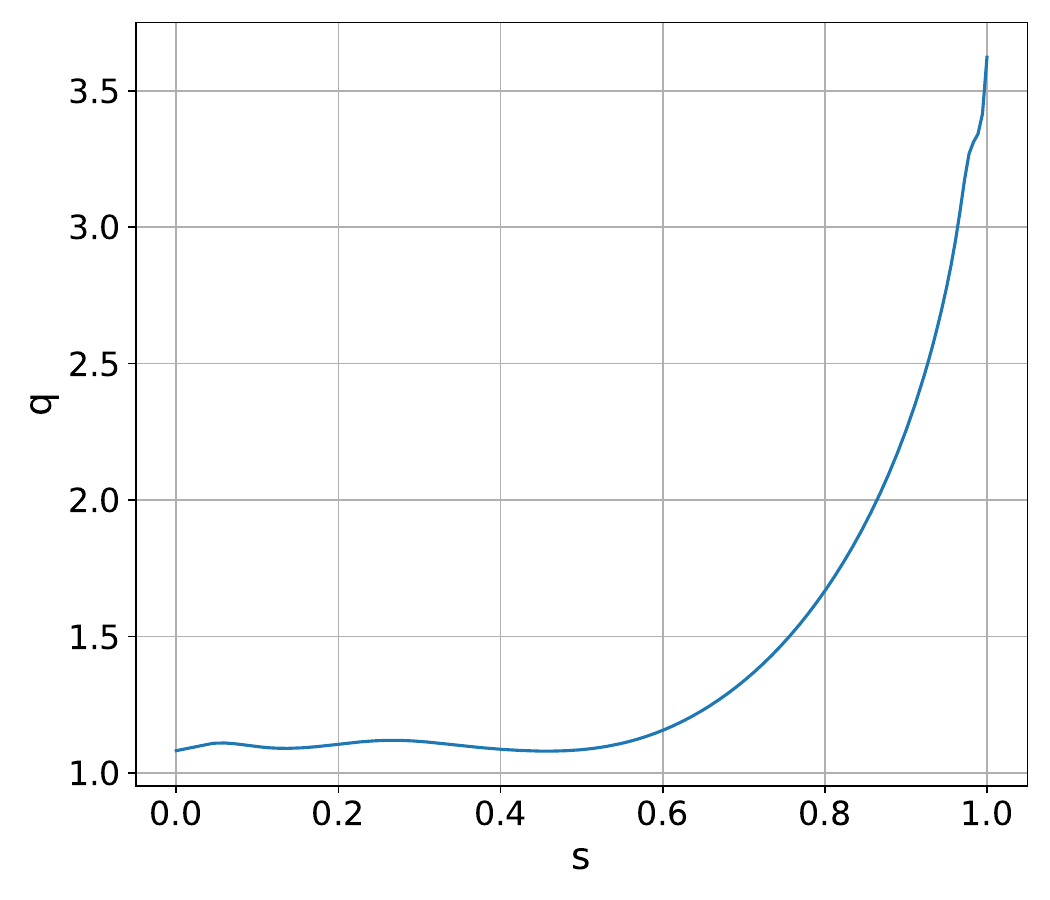}
\label{fig:q_ITER}
\end{minipage}
\hfill
\begin{minipage}[h]{0.49\linewidth}
\includegraphics[width=1\linewidth]{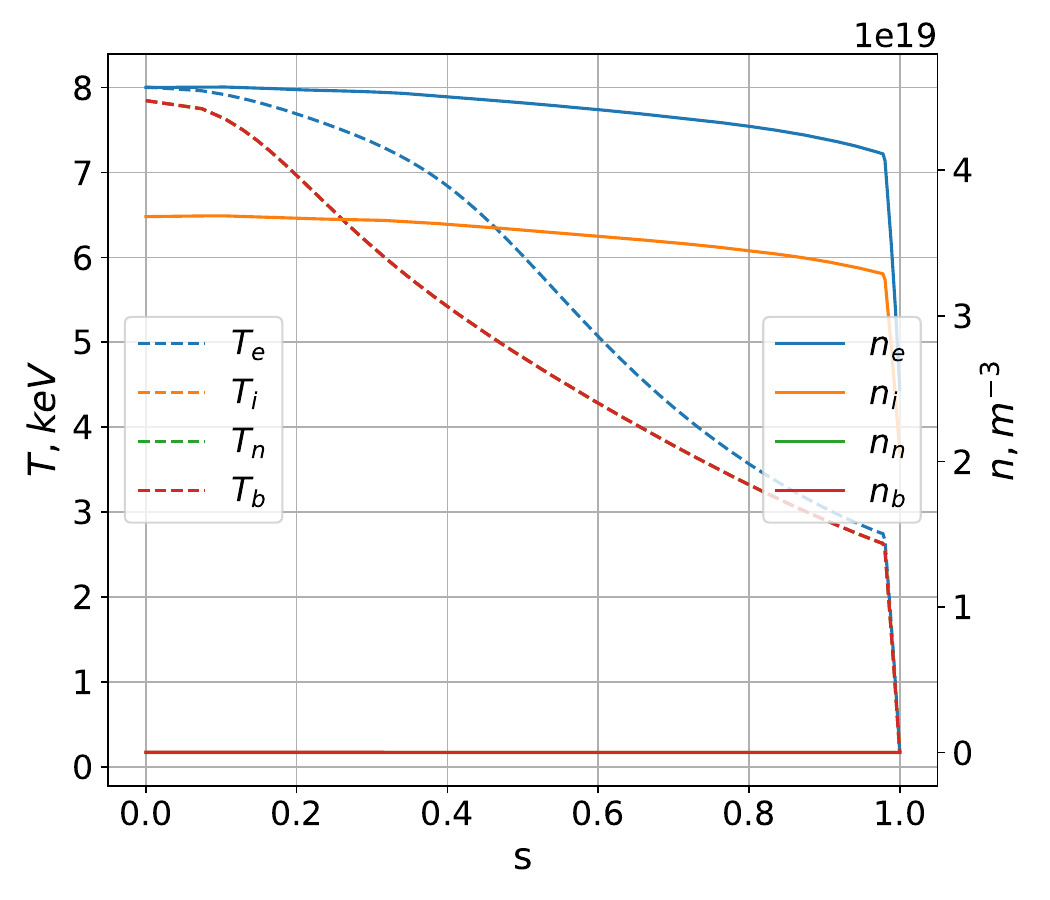}

\end{minipage}
\end{center}    
\caption{PFPO ITER case \cite{Hayward2022}: a) - Safety factor $q$ profile; b) - temperature and density profiles}
\label{fig:nT_ITER}
\end{figure}

\begin{figure}[h]
\begin{center}
\begin{minipage}[h]{0.49\linewidth}
\includegraphics[width=1\linewidth]{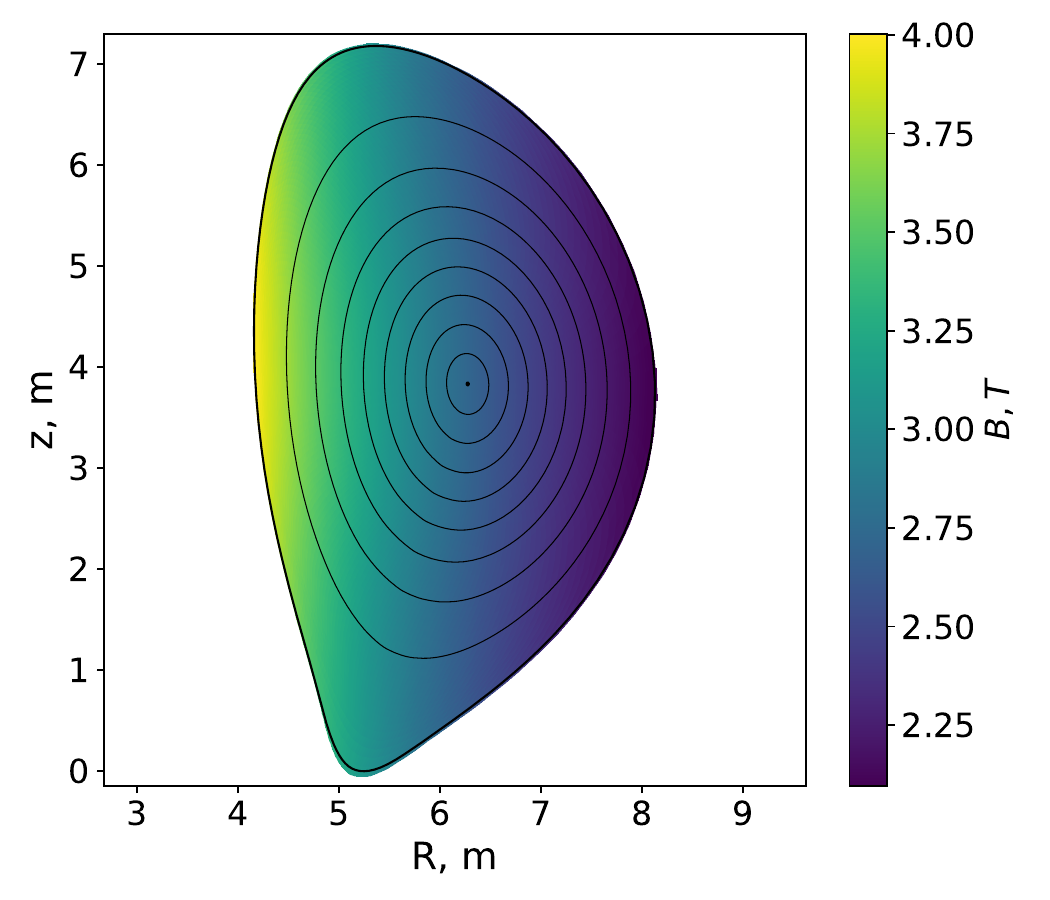}
\end{minipage}
\hfill
\begin{minipage}[h]{0.49\linewidth}
\includegraphics[width=1\linewidth]{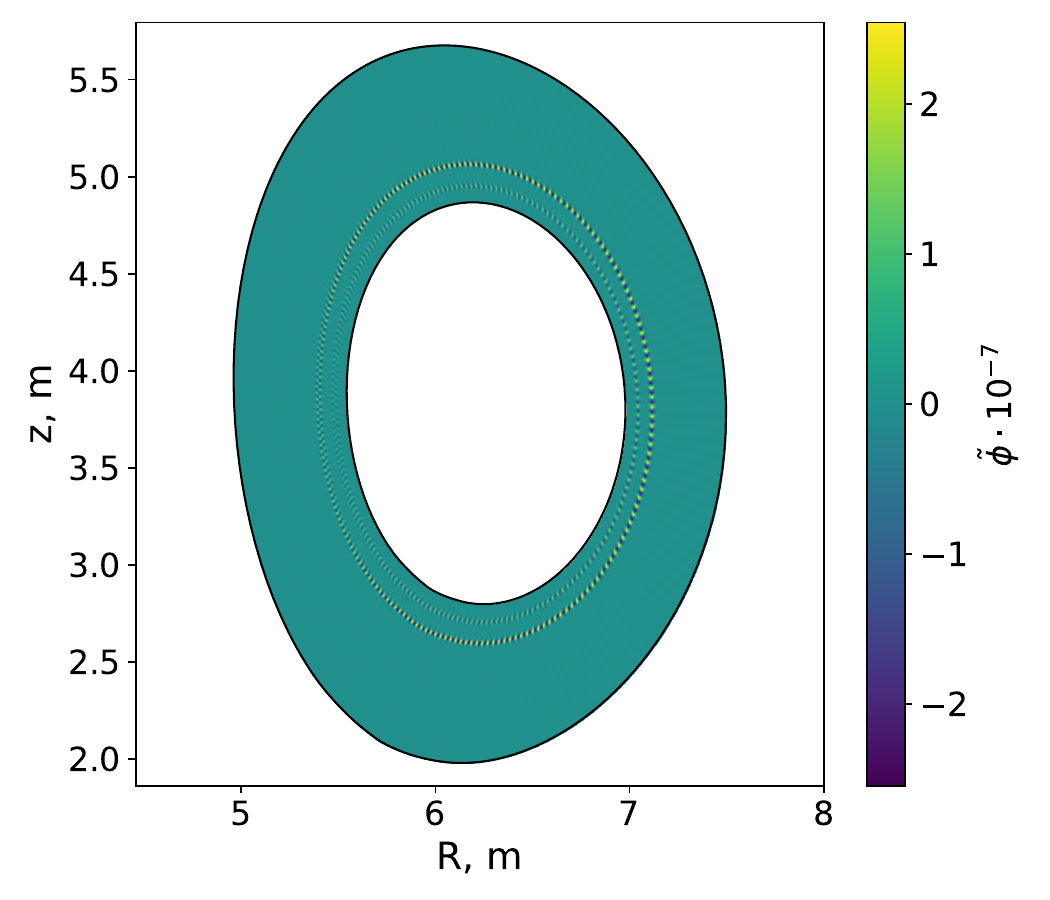}

\end{minipage}
\end{center}    
\caption{ITER PFPO scenario space structure \cite{Hayward2022}: a) - Equilibrium magnetic field; b) - Scalar potential : two ITGs are observed: dominant one at $s=0.47$ and one which is one magnitude weaker at $s=0.43$}
\label{fig:phi_ITER}
\end{figure}

Temperature, density and q profiles are consistent with with the magnetic equilibrium \cite{Rofman2026} and shown in Fig. \ref{fig:nT_ITER}. On-axis electron and hydrogen temperatures are taken to be $T_e(s=0)=T_i(s=0)\approx 7.85\,keV$ with the top pedestal values of $T_e\approx 3\,keV$. On-axis electron density is set to be $n_e(s=0)=4.26\cdot10^{19}\,m^{-3}$. The magnetic equilibrium (Fig. \ref{fig:phi_ITER}) was taken from the same paper \cite{Hayward2022} which was produced by running the CHEASE MHD equilibrium code. ITG was excited on the single toroidal mode number again, but with the value of $n=180$. This time there were observed two ITG modes at different radii, as shown in the Fig. \ref{fig:phi_ITER}: dominant one at $s=0.47$ and one which is one magnitude weaker at $s=0.43$ and a TEM mode. All the following results are presented only for the dominant ITG. All simulations were performed on the spatial grid: $N_s =384, N_\theta=1472, N_\phi=736$ with $6.4\cdot10^{7}$ number of markers for background and fast hydrogen ions and electrons and $1.6\cdot10^{7}$ for impurities. Length of the simulations was $10^{5}\cdot\Omega_{ci}$ with a time step of $20\cdot\Omega_{ci}$.
\subsubsection{Results of electrostatic simulations} \label{sec:ITER_es}
First set of simulations was performed with 3 species: adiabatic electrons, thermal hydrogen, Maxwellian hydrogen EPs were taken, omitting the impurities (beryllium and neon). Electrostatic simulations with adiabatic electrons have been considered in this first set of simulations, for a more consistent comparison with the case presented in the previous section. Temperature profile of the fast particles for simplicity was taken to be flat. Concentrations of the EPs were taken $n_f/n_{i}=10\%$ and $n_f/n_{i}=1\%$ as a more realistic one.

Resulting temperature scan is shown in the Fig. \ref{fig:gamma_ITER}. For both densities the trend is similar to what was observed in the previous chapter in the destabilizing regime. Indeed, the temperature here was flat, so $\eta_f$ parameter is equal to $0$, corresponding to the EP destabilization of the mode. Additionally, the gamma values at the temperature of $100$ are consistent with the DE: relative growth rate reductions are $0.1$ and $0.01$ for $10\%$ and $1\%$ concentrations. Note that, for the concentration of $1\%$, the linear growth rate modification is of the same order of the error bar, therefore we can conclude that the ITG stabilization is negligible with respect to the case of $10\%$.

\subsubsection{Results of electromagnetic simulations}
\begin{figure}[h]
\begin{center}
\begin{minipage}[h]{0.49\linewidth}
\includegraphics[width=1\linewidth]{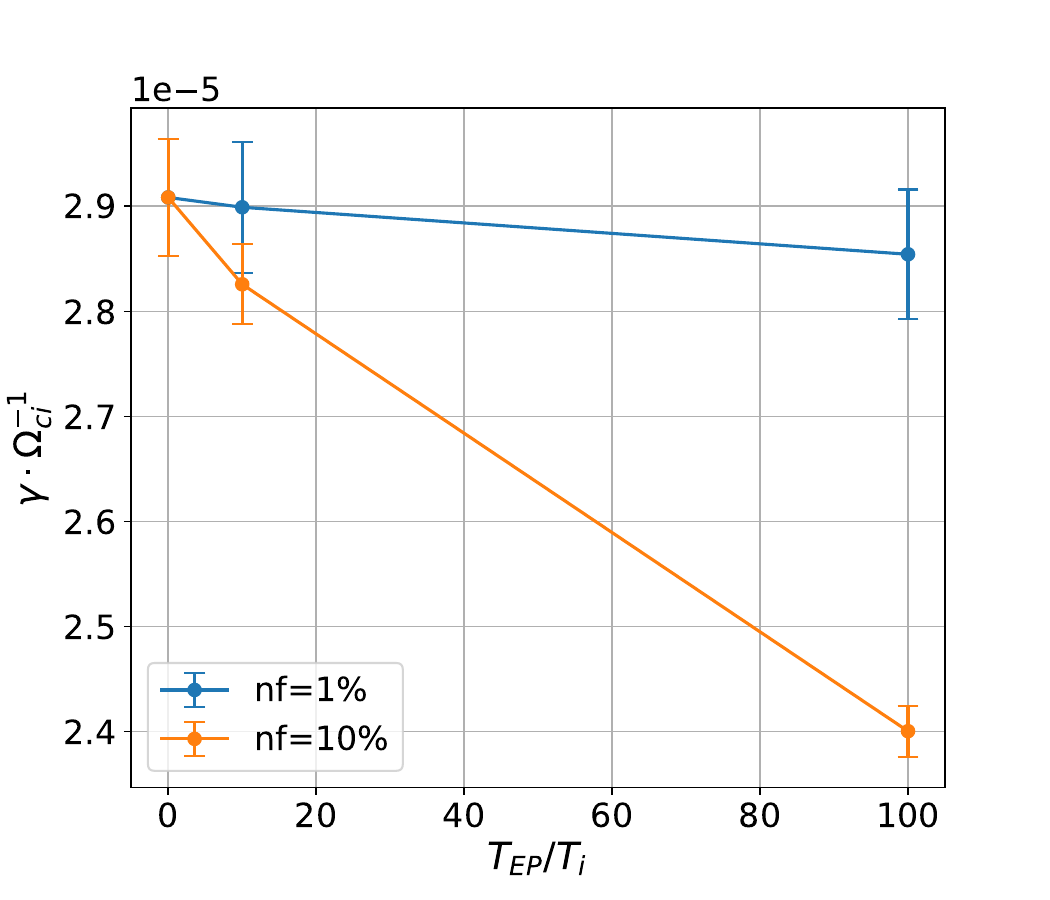}
\end{minipage}
\hfill
\begin{minipage}[h]{0.49\linewidth}
\includegraphics[width=1\linewidth]{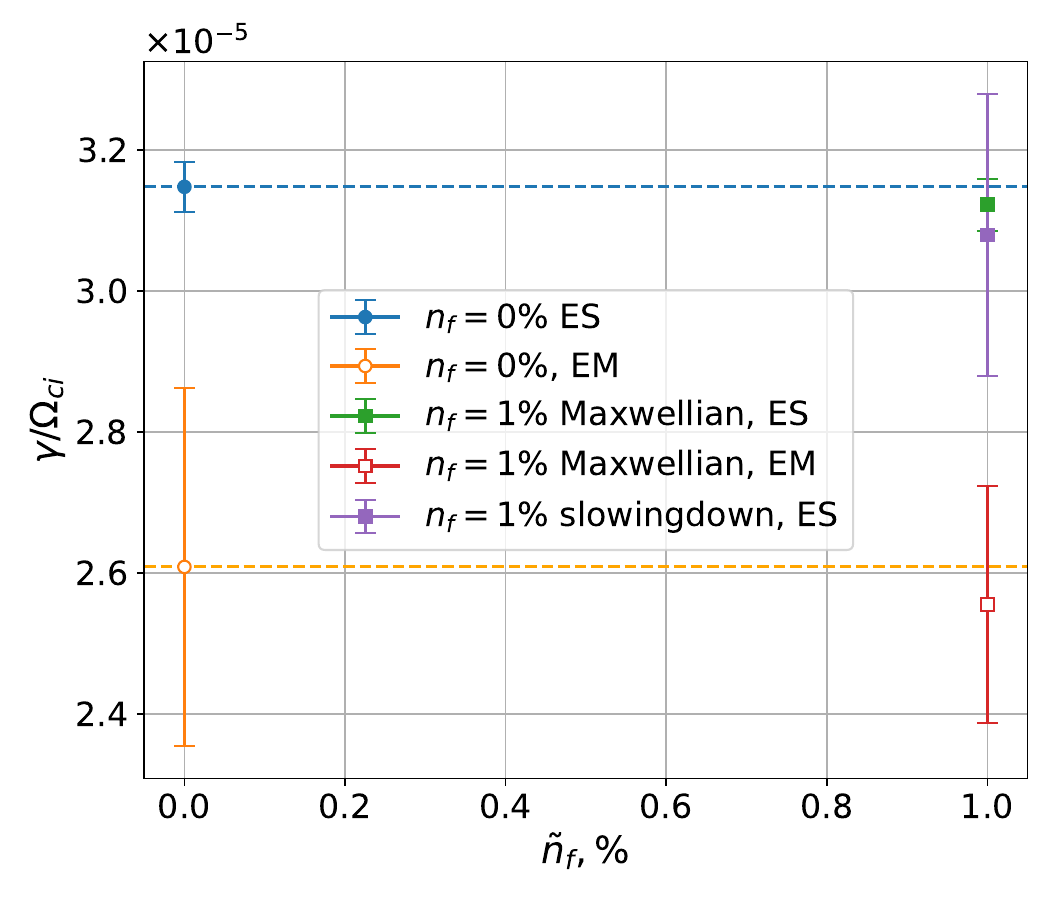}
\end{minipage}
\end{center}
\caption{a) - growth rate $\gamma(T_f)$ scan in EP temperature for constant EP densities of $1\%$ and $10\%$ in ITER case \cite{Hayward2022} in destabilizing regime with $\eta_f=0$: decreasing trend is confirmed as for the simplified case, dilution effect is observed at $10\%$ density, while at $1\%$ it is inside the error-bars; b) - growth rate for different simulations with and without $1\%$ of EPs: dilution effect is at least 3 times smaller than the $\beta$ stabilization. EP distribution has no influence on the growth rate}
\label{fig:gamma_ITER}
\end{figure}

Second set of simulations was performed with and without EPs (both Maxwellian and SD with $\xi=-0.66$ and $\sigma=0.22$ - values estimated from the paper \cite{Brochard2025}) with concentration of $1\%$ with a temperature equivalent to the temperature of the SD EPs from the hydrogen NBI of $870\,keV$. It corresponds to the parameter $T_f/T_i\approx30$. Simulations were performed in the presence of neon and beryllium impurities with the same parameters as in \cite{Hayward2022}. Electromagnetic and electrostatic simulations with adiabatic electrons were performed. Electron mass was increased $m_H/m_e=400$ compared to the realistic one, in order to reduce the numerical cost. Growth rates are shown in Fig. \ref{fig:gamma_ITER}. In the latter, dilution effect is again present with the relative reduction of $3\%$ which is larger than the concentration dilution, due to two reasons: in $T_f/T_i\approx30$ regime there is a remaining interaction between the ITG and the particles and the error-bars are again too large to consider values to be exact. 
At the same time, electromagnetic simulations show larger $\beta_f$ stabilization \cite{Weiland1992}: the reduction of growth rate values without the EP going from electrostatic to electromagnetic is $19\%$ which is at least 3 times larger than the EP stabilization in both cases (in electromagnetic case EP stabilization is $6\%$ again inside the error-bars). 
Note that these results are consistent with \cite{Rofman2026}.

\section{Conclusions} \label{conclusion}
This article investigates the dynamics of ion temperature gradient modes, in burning plasma regime, in the presence of energetic particles. This problem is considered as a high-relevance topic in the fusion community, and, at different levels, it has been considered earlier in a number of highly-cited publications. Results of these publications are being included into modern designs of the fusion plasma devices (reactors such as SPARC \cite{Creely2020} in the USA or BEST in China). However, the earlier work often relies on a number of simplifications which  may be incompatible with the reactor plasmas: in particular, in some papers, the local flux-tube approach is used, which has been demonstrated to be outside the domain of validity; the ratio of the fast-ion energy to the bulk-plasma temperature is significantly lower than in a reactor; a Maxwellian distribution function is used for the energetic ions. In our paper, we investigate how these simplifications may affect the physics of the energetic-particle interaction with ion temperature gradient modes. Results of our investigation may have implications for future reactor designs. 

In this work the linear interaction between the energetic particles (EPs) and ion temperature gradient modes (ITG) was studied with the help of the ORB5 gyrokinetic code with the main focus in the burning plasma regime and realistic plasma conditions. Two main stabilization mechanisms in the electrostatic simulations were observed: dilution effect (DE) and direct dispersion relation modification (DDRM). First dominates at high EP temperatures when they weakly interact with the mode, while the second is mostly seen in the intermediate temperature region and is a direct effect of modification of the mode dispersion relation.   

In a simplified electrostatic case \cite{Mishchenko2022}, these effects were observed for a Maxwellian EP distribution, consistent with the findings of \cite{DiSiena2018}, precisely: in stabilization regime the DDRM was observed at $T_f/T_i \approx 10$ and DE at $T_f/T_i \approx 50$, while in the destabilization regime the growth rate was monotonically decreasing to the dilution value with $T_f$. In Maxwellian case the main parameter that controls the regime of stabilization or destabilization of the ITG by EPs was confirmed to be $\eta_f^{-1}$ which leads to the stabilization at $\eta_f^{-1}\le 1.5$. As a complementary study, for constant $\beta_f$ conditions that are relevant to experimental plasma scenarios, by varying the EP temperature the stabilization minimum was shifted toward lower temperatures due to increased EP density, confirming the interplay between concentration and direct interaction effects.

Research was expanded to the case of more realistic EP distributions, namely anisotropic and isotropic slowing-down, that are good approximations to the neutral beam injected (NBI) EPs and fusion born EPs respectively. For both distributions the monotonic decrease of the growth rate with temperature was observed, showing inefficient DDRM. It was further confirmed with the mode particle resonance diagnostic (MPR) that shows effective power exchange in the parts of the phase space where particles are the most distributed but not where they fulfill the DDRM condition. The specific criterion similar to the case of the Maxwellian distribution was derived for the isotropic case confirming the simulation results, showing that at the low temperatures EP have stabilizing contribution, while passing the values of $T_f/T_i \approx 4$ switch to the destabilization.

Simplified analytical case studies were further expanded to the experimental ITER pre-fusion operation scenario of \cite{Hayward2022}. For this case same behavior was observed: decreasing trend of the growth rate was seen for unrealistically increased densities $n_f=10\%$, while for real values the effect was proportionally decreased and appeared within the error-bars. Electromagnetic simulations showed a significantly stronger $\beta$-induced stabilization (three times larger than EP-related effects).

Overall, the results confirm the impact of the EP on the ITG modes in linear regime and show that it strongly depends on the various parameters such as the EP distribution in velocity space and density and temperature gradients. Maxwellian EPs can provide significant stabilization by the DDRM mechanism for intermediate temperatures and specific density and temperature profiles while slowing-down EPs both of NBI and fusion origin can only be responsible for the DE stabilization.
In summary, the results of this paper show that in the present tokamak scenario, a combination of dilution and DDRM is mostly present, when considering electrostatic simulation with adiabatic electrons. But if one wishes to study more realistic burning plasma scenarios, like in the case of ITER considered here, beta stabilization is dominant and cannot be neglected, if one wants to have experimentally relevant predictions. The impact of the EP on the ITG modes, will be devoted to the electromagnetic studies of the EP ITG interactions for both linear and non-linear regimes of the ITG in a simplified analytical and experimental ITER cases including relevant particle distributions.

\section{Acknowledgements} \label{acknowledgements}
This work was performed with the ORB5 code on the Leonardo and Pitagora Supercomputers at CINECA. The authors expresses gratitude for the ORB5 team and especially to Philipp Lauber for valuable advises, help in launching simulations and comments regarding physics during ORB5 team meetings. The authors wish to thank Ivan Novikau for valuable help in understanding of the MPR diagnostics, Pierre Morel for help in analytical derivations and Laurent Villard and Fulvio Zonca for discussion of the results and advises for choosing the topics of studies.

This work has been carried out within the framework of the PEPR Suprafusion project.
\addtocontents{toc}{\protect\setcounter{tocdepth}{2}} 

\section{References} \label{sec:references}
\printbibliography[heading=none]
\end{document}